\documentclass{aa} 
\usepackage{graphicx}
\usepackage{txfonts}
\begin{document}
   \title{A Catalogue of  Be Stars in the Direction of the Galactic Bulge}

   \author{B.E. Sabogal
          \inst{1}
          \and
          R.E. Mennickent
	  \inst{1}
	  \and G. Pietrzy\'nski
	  \inst{1,2}
	  \and J.A. Garc\'{\i}a
	  \inst{1}
	  \and W. Gieren
	  \inst{1}
	  \and  Z. Kolaczkowski
	  \inst{1} 
          }

   \offprints{B.E. Sabogal, e-mail: bsabogal@udec.cl}

   \institute{Universidad de Concepci\'on, Departamento de F\'{\i}sica,
      Casilla 160-C, Concepci\'on, Chile\\
 %                   \email{bsabogal@udec.cl}    
         \and
             Warsaw University Observatory, Al. Ujazdowskie 4,00-478, Warsaw, Poland\\
             }

  \abstract
   {Detailed studies of Be stars in environments with different metallicities like the Magellanic Clouds or the Galactic bulge are necessary to understand the formation and evolution mechanisms of the circumstellar disks. However, a detailed study of Be stars in the direction of the bulge of our own galaxy has not been performed until now.}
   {The aim of this work is to report  the first systematic  search for Be star candidates in the direction of the Galactic
Bulge. We present the full catalogue, give a brief description of the stellar variability seen, and show some light curve examples.}
   {We searched for stars matching specific criteria of magnitude, color and
variability in the $I$ band. Our search was conducted on the 48 OGLE II fields of the Galactic Bulge. }
   {This search has resulted in 29053 Be star candidates, 198 of them showing periodic light variations. Nearly 1500 stars  in this final sample are almost certainly Be stars, providing an ideal sample for spectroscopic multiobject follow-up studies. }
   {}


   \maketitle
%
%________________________________________________________________

\section{Introduction}
 Be stars are non-supergiant fast rotator B stars whose spectra have, or had at some time, one or more Balmer lines in emission (Collins 1987).  This  emission is originated from a flattened circumstellar  disk and  can come and go episodically on time scales of days to decades.  The responsible mechanisms for the production and dynamics of the circumstellar gas are still not constrained. Possible mechanisms
include non-radial pulsations, wind-compressed disk model,  magnetic activity and binarity (Porter $\&$ Rivinius 2003, and  references therein).\\
Be stars are variable in brightness on three time scales that are often superimposed. Many of them (especially early type Be stars) show short-term photometric variability on time-scales of 0.2 to 2 days and amplitudes up 0.1 magnitudes, caused by non-radial pulsation or rotation (Percy et al. 2002, 2004). Some have mid-term variations on times scales form weeks to months, probably due to density waves within the disk (Sterken et al. 1996). Their amplitudes go to up to 0.2 magnitudes. They show also long-term variations  from years to decades, with amplitudes up 0.8 magnitudes (Mennickent, Vogt, \& Sterken 1994; Pavlovski et al. 1997; Hubert  $\&$ Floquet 1998; Percy $\&$ Bakos 2001). Stagg (1987) found that this type of variability occurres in at least half of the Be stars. A few of Be stars are close binaries  and other present ejection process due to magnetic activity resulting in outbursts (Hubert et al. 1997).\\
Many Galactic Be stars have been surveyed for photometric variability in order to detect and confirm short-term or long-term variations and to find correlations between them and get clues on the physical processes in Be stars. For instance, Hubert $\&$ Floquet (1998) investigated the short-period variability of Be stars using analysis based on the
Fourier and CLEAN algorithms on the Hipparcos photometry. Percy et al. (2002, 2004) analyzed a large sample of stars with Hipparcos photometry using a form of autocorrelation function. Typical problems in these studies are the gaps in the time distribution of the measurements and the limitations of the used algorithms.\\ 
In recent years, many Be-star like variables have been discovered in the Magellanic Clouds, showing a big variety of light curves, some of them reminding those of the Galactic Be stars and others never observed in that type of stars. Keller et al. (2002) concluded that most of these blue variables should be Be stars. Searches for Be stars in the Magellanic Clouds
were performed by Mennickent et al. (2002), Keller et al. (1999, 2002) and Sabogal et al. (2005) on the basis of selection criteria applied to different photometric databases (OGLE-II and MACHO),  and took into account amplitude of variability
and ranges of color-magnitudes in the selection process.   
De Wit et al. (2006) investigated a subsample of the blue variables found by Mennickent et al. (2002) in the Small Magellanic Cloud and found that the photometric variability of these Be stars is due to variations in the amount of Brehmstrahlung due to the evolution of the circumstellar gas from a disk-shaped envelope towards a ring-like structure. \\
The study of Be stars  is relevant to make contributions to several important branches of stellar physics. In particular, detailed studies of Be stars in environments with different metallicities like the Magellanic Clouds or the Galactic bulge is crucial to understand the formation and evolution mechanisms of the circumstellar disks. However, a detailed study of Be stars in the direction of the bulge of our own galaxy has not been performed until now.\\
A very large number of stars was observed in the region of the Galactic Bulge during the course of the second phase of the Optical Gravitational Lensing Experiment (OGLE II) (Udalski, Kubiak \& Szymanski 1997). We have performed a search for Be star candidates into this database. Here we present the  results of this search.

%__________________________________________________________________

\section{The Data}
During  the course of the OGLE II project (Udalski, Kubiak  \& Szymanski 1997; Udalski et al. 2002), VI photometry maps of the Galactic center region were obtained.  They contain astrometrical and photometrical data of about 30 million stars in  49 observed fields, which are already publically available in the OGLE web page. The time base of these OGLE II observations was three years,  and the 49 fields were monitored on every night with good seeing conditions. The majority of observations were obtained using the I-band filter and a smaller number of them was obtained through the V-band filter  (Udalski et al. 2002). From these maps we looked for stars with absolute V-band  magnitudes in the typical range of the Galactic Be stars, i.e. $-6 < M_V < 0$ (Wegner 2000; Garmany \& Humphreys 1985). Assuming a distance modulus for the Galactic Bulge of 14.5 mag (Mc Namara et al. 2000), we obtained apparent magnitudes in the range of  $8.5 + A_V < V < 14.5 + A_V$, where $A_V$ is mean extinction value for each field of the Galactic Bulge obtained from Sumi (2004).  As the saturation limit of the detector used by OGLE II project is 12 mag in the V-band our search was restricted to the range $12 < V < 14.5 + A_V$ when the inferior limit $8.5 + A_V$ was less than 12 mag.\\
We also constrained our search to the possible range of colours $V-I$ of classical Be stars. These colours are reddened due to the circumstellar disks and interstellar reddening (Wisniewski \& Bjorkman, 2006), which is very strong in the direction of the galactic center. Therefore, we searched for stars with $-0.35  < V-I < 0.8 + E(V-I)$, where $E(V-I)$ is mean reddening value for each field of the Galactic Bulge obtained also from Sumi (2004). He obtained  extinction and reddening maps in the V and I bands for 48 OGLE II fields of the Galactic center region by using the average value of  the ratio of  total to selective extinction $R_{VI}= A_V/E(V-I)= 1.964 \pm 0.085$. Because of the variable extinction in the direction of the Galactic center, these  mean extinction and mean reddening values cover the following  ranges: $0.676 < E(V-I) < 2.918$  mag, and  $1.327 < A_V < 5.733$ mag. It is important to note that only in some fields  extinction and reddening reach high values. In particular,  the OGLE-II fields $BUL-SC5, BUL-SC37$ and $BUL-SC43$ have the highest values of these parameters. Field $BUL-SC44$ was not used in this search for Be star candidates because it was discarded by Sumi (2004) for his measurements of extinction and  $E(V-I)$ and $A_V$ are not reported for this field.\\
It is important to note that when looking to the galactic bulge we observe many
foreground and background stars. For this reason our selected sample might include stars 
whose distance moduli is different from that required by our selection criteria.
In practice, this implies that the selected sample could result contaminated by 
foreground stars dimmer than usual Be stars and background stars brighter than normal Be stars.\\
It is worth noting that although OGLE II Galactic Bulge fields have high stellar densities and very strong and variable extinction, accurate photometry could be obtained from them (0.02-0.04 mag accuracy). However, $I$ magnitudes and $V-I$ colours for very red stars can be different from the standard values, reaching brighter $I$ magnitudes (up 0.25 mag) and redder $V-I$ values for stars with $(V-I) > 2$ (Udalski et al. 2002).\\
Another important aspect of OGLE II Galactic Bulge fields is that due to crowding many stars can be blended, in particular those with I values close to 18 mag and fainter (Sumi et al. 2004). \\
Keeping in mind the above limitations a total of 173404 stars were selected within the expected ranges of magnitudes and colours of Be stars at the Galactic Bulge distance. We have called them {\it{Be star precandidates}}. Table 1 shows for each field of the Galactic Bulge the total number of stars in the  OGLE II database and the total number of stars with $V$ magnitudes and $V-I$ colours similar to those of Be stars. It also shows the number of selected Be star candidates and the number of most certain Be star candidates, whose selection criteria are described in the following section. Finally this table shows the total number of stars for each column and its percentage respect to the total sample of stars in the direction of the Galactic bulge in the OGLE-II database. 

%                                             Two column Table 
%_____________________________________________________________

\begin{table*}
 \caption{Census of Be Star Precandidates and Candidates in the Direction of the Galactic Bulge.}
 \label{1}
 \centering 
 \begin{tabular}{@{}ccccccccc}
  \hline
  \hline
Field & & Total & & Total of Be star & & Total of Be & & Total of most certain Be \\
 & & of stars & & precandidates & & star candidates & & star candidates \\
  \hline
$BUL-SC1$ & & 729852 & & 2407 & & 273 & & 30 \\
$BUL-SC2$ & & 803269 & & 2501 & & 460 & & 22 \\
$BUL-SC3$ & & 805587 & & 9694 & & 1756 & & 121 \\
$BUL-SC4$ & & 774091 & & 7449 & & 1433 & & 95 \\
$BUL-SC5$ & & 433990 & & 7306 & & 1194 & & 70 \\
$BUL-SC6$ & & 514084 & & 1055 & & 209 & & 10 \\
$BUL-SC7$ & & 462748 & & 1062 & & 260 & & 11 \\
$BUL-SC8$ & & 401813 & & 1833 & & 358 & & 8 \\
$BUL-SC9$ & & 330338 & & 1865 & & 291 & & 10 \\
$BUL-SC10$ & & 458816 & & 4112 & & 320 & & 31 \\
$BUL-SC11$ & & 425984 & & 3562 & & 313 & & 36 \\
$BUL-SC12$ & & 534720 & & 4282 & & 700 & & 40 \\
$BUL-SC13$ & & 569850 & & 3877 & & 644 & & 31 \\
$BUL-SC14$ & & 619028 & & 1735 & & 274 & & 10 \\
$BUL-SC15$ & & 600787 & & 1859 & & 380 & & 10 \\
$BUL-SC16$ & & 699804 & & 3053 & & 599 & & 27 \\
$BUL-SC17$ & & 687019 & & 2387 & & 460 & & 17 \\
$BUL-SC18$ & & 749265 & & 3511 & & 599 & & 21 \\
$BUL-SC19$ & & 732089 & & 3664 & & 617 & & 20 \\
$BUL-SC20$ & & 785317 & & 4566 & & 847 & & 20 \\
$BUL-SC21$ & & 882518 & & 3611 & & 640 & & 30 \\
$BUL-SC22$ & & 715301 & & 6434 & & 1290 & & 51 \\
$BUL-SC23$ & & 723687 & & 4697 & & 890 & & 48 \\
$BUL-SC24$ & & 612189 & & 2362 & & 399 & & 18 \\
$BUL-SC25$ & & 622326 & & 2224 & & 397 & & 15 \\
$BUL-SC26$ & & 728200 & & 1994 & &  216 & & 19 \\
$BUL-SC27$ & & 690785 & & 1890 & &  280 & & 11 \\
$BUL-SC28$ & & 405799 & & 1135 & &  255 & & 19 \\
$BUL-SC29$ & & 491941 & & 1098 & &  176 & & 7 \\
$BUL-SC30$ & & 762481 & & 3731 & &  793 & & 24 \\
$BUL-SC31$ & & 790471 & & 3673 & &  590 & & 28 \\
$BUL-SC32$ & & 797493 & & 2902 & &  469 & & 22 \\
$BUL-SC33$ & & 738508 & & 2502 & &  566 & & 25 \\
$BUL-SC34$ & & 960656 & & 4967 & &  1144 & & 57 \\
$BUL-SC35$ & & 770940 & & 3570 & &  756 & & 29 \\
$BUL-SC36$ & & 873472 & & 3005 & &  561 & & 37 \\
$BUL-SC37$ & & 664424 & & 10180 & &  1740 & & 80 \\
$BUL-SC38$ & & 710234 & & 2620 & &  507 & & 37 \\
$BUL-SC39$ & & 584316 & & 5897 & &  1020 & & 44 \\
$BUL-SC40$ & & 630774 & & 2795 & &  493 & & 24 \\
$BUL-SC41$ & & 603404 & & 2618 & &  614 & & 28 \\
$BUL-SC42$ & & 600519 & & 4367 & &  1024 & & 47 \\
$BUL-SC43$ & & 474367 & & 1406 & &  242 & & 17 \\
$BUL-SC45$ & & 627412 & & 2326 & &  525 & & 9 \\
$BUL-SC46$ & & 551815 & & 2078 & &  374 & & 27 \\
$BUL-SC47$ & & 300705 & & 3917 & &  517 & & 30 \\
$BUL-SC48$ & & 286907 & & 5137 & &  352 & & 41 \\
$BUL-SC49$ & & 251629 & & 2134 & &  236 & & 24 \\
Total of stars & & $\sim$ $30 \times 10^6$ & & 173404 & & 29053 & &  1448 \\
Percentage & & 100 \% & & 0.6 \% & & 0.1 \% & & $5 \times 10^{-3}$ \% \\
  \hline
 \end{tabular}
\end{table*}

\section{Results}
In order to obtain a list of certain Be star candidates in the direction of the bulge of our Galaxy, the following selection criteria were applied on the sample previously obtained (Be star precandidates).

\subsection{Selection Criteria}
In order to determine the global variability properties of our sample and decide about the method of selection of Be star candidates, five fields: $BUL-SC1, BUL-SC10, BUL-SC11, BUL-SC26$ and  $BUL-SC48$, far and near to the central region of the Galactic Bulge, were selected as test fields. Table 2 shows equatorial and galactic central coordinates of the selected fields.  The I-band light curves for each previously selected star of  the five fields  were extracted from the OGLE II database.\\
Taking on account that  large standard deviations of the I-band magnitudes are usually good indicators of stellar variability, a robust estimate of its mean magnitude $\overline{I_R}$ and standard deviation $\sigma_R$ was obtained for each star on the mentioned fields. This step was achieved by using  a statistical C-program which is less sensitive to outliers than the usual ones (the C-code called SIGCOL is freely available at
http://www.spaennare.se/ssphot.html. The reader can find at this page the details about the methode used by the program to calculate the robust estimate of each variable). Then the stars were grouped in 25 magnitude bins  and the average $\sigma_R$  and the average $\overline{I_R}$  were calculated per bin. In order  to obtain a function $\Sigma(\overline{I_R})$ that represents $\sigma_R$ as a function of $\overline{I_R}$, the data were fitted with a third-order polynomial.  Stars with $\sigma_R$  values smaller than  $\alpha \Sigma(\overline{I_R})$  were rejected  ($\alpha$ is the selected threshold of variability).  To select the  $\alpha$ value, we plotted  $\sigma_R$ against $\overline{I_R}$ for all stars on each field, and the obtained polynomial $\Sigma(\overline{I_R})$. Figure 1 shows an example of this graph for field $BUL-SC11$.  Solid line represents the fitted polynomial.  It is seen on this figure  that the distribution of the majority of stars  is well fitted by this polynomial and that the added function $\alpha \Sigma(\overline{I_R})$ (dashed line)  is a good limit between this distribution and that of possible variable stars.  The best  $\alpha$ value  was  1.6 for  for field $BUL-SC1$, 1.5 for $BUL-SC10$, 1.6 for $BUL-SC11$, 1.7   for $BUL-SC26$ and 2.0 for $BUL-SC48$.  For this reason we selected  $\alpha$ = 1.5 as a conservative value for these five fields and used the same value for the remaining 43 fields.\\
\begin{figure}
\centering
\scalebox{.9}[.9]{\includegraphics[angle=0,width=10cm]{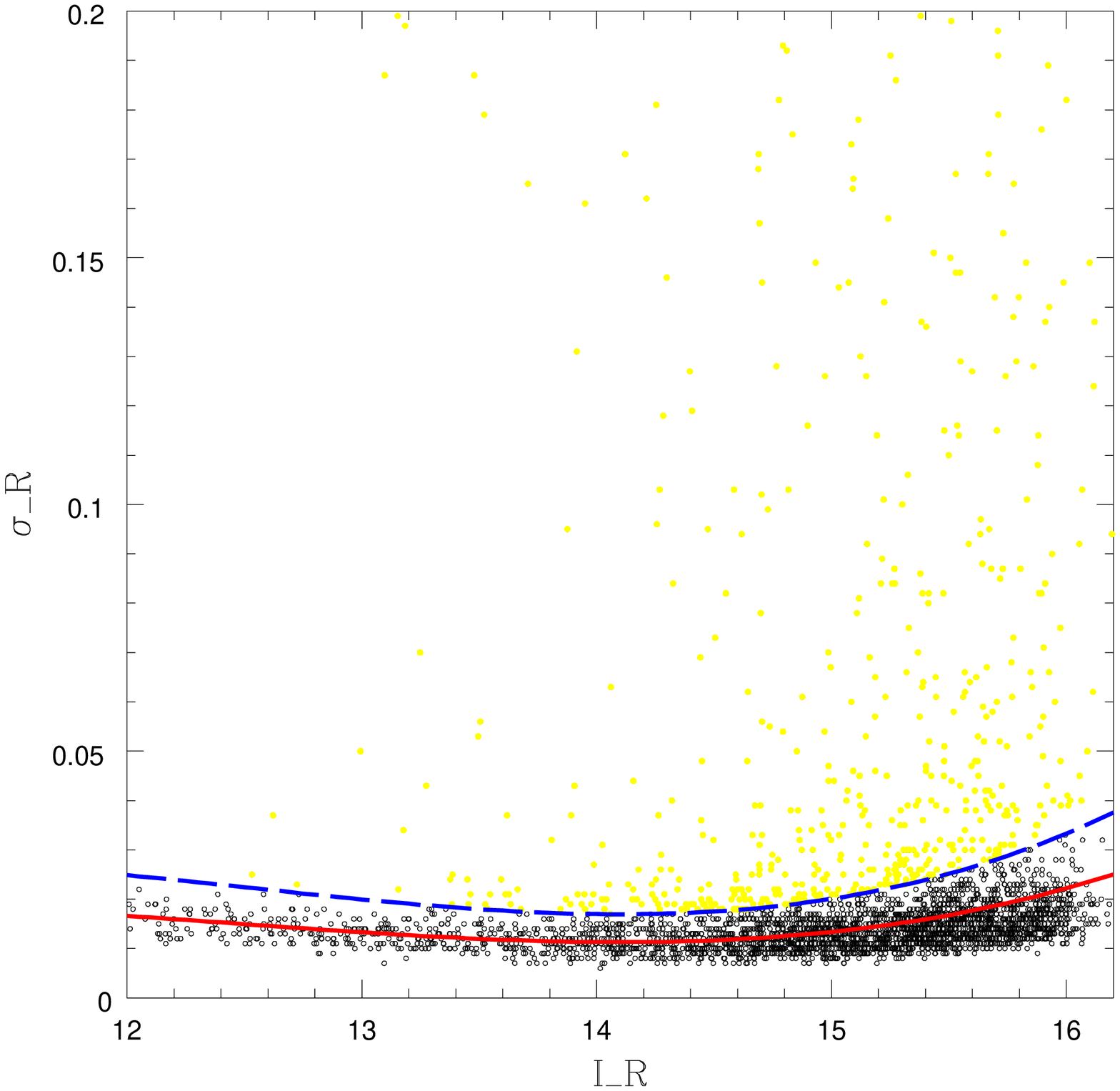}}
 \caption{Standar deviation $\sigma_R$ versus mean I-band magnitude $\overline{I_R}$ for all Be star precandidates of $BUL-SC11$ field. Solid line is the third-order polynomial $\Sigma(\overline{I_R})$ fitted to the data as is explained on the text. Dashed line is the $\alpha \Sigma(\overline{I_R})$ funtion, with   $\alpha = 1.5$.}
  \label{1}
\end{figure}

In order to clean many light curves that showed sets of points much brighter or much fainter in magnitudes than those of the star (figure 2 presents the time series of two stars of $BUL-SC39$ field showing this problem) we constructed  a pipeline that selects data between $\overline{I_R} - 3\sigma_R$ and $\overline{I_R} + 3\sigma_R$ for each star. Then the pipeline calculates  amplitudes $A_I$ for the clean light curves, by using the equation $A_I = MaxI - minI$, where MaxI and minI are the maximum and minimum values of the I magnitudes.  These amplitudes are  always  less than 1.00 mag.  Finally,  we  select  stars with amplitudes in the typical range of classical Be stars and also other type of blue variables like those found in the Magellanic Clouds (Mennickent et al. 2002). This step was achieved rejecting stars with amplitudes in the I-band less than 0.05 mag and greater than 1.00 mag. As a test of the superior limit of 1.00 mag we calculated amplitudes for the uncleaned light curves of all stars  in several fields,  and visually inspected the light curves of those stars with amplitudes greater than 1.00. We found that all of them were similar to those of figure 2.  It is worthy to note that about 30 \% of the light curves of every fields presented photometric contamination like that shown in figure 2. The reason of this problem is yet in study (Pietrzy\'nski 2004).\\
\begin{figure}
\centering
\scalebox{.9}[.9]{\includegraphics[angle=0,width=10cm]{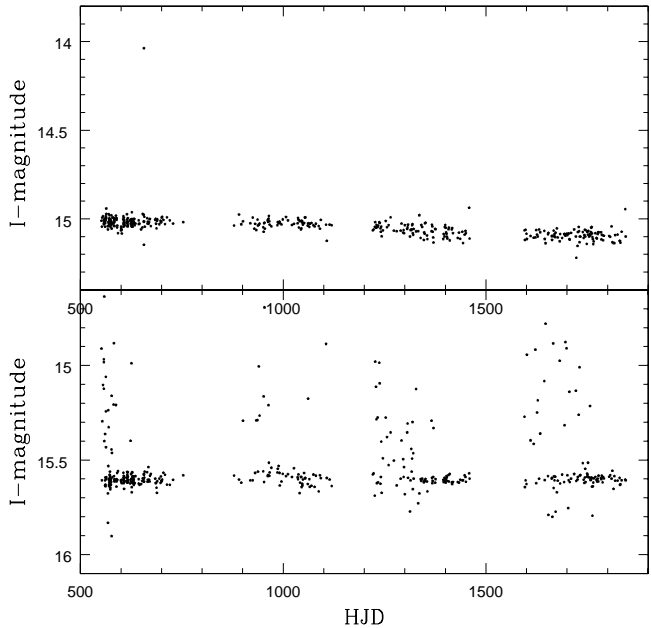}}
 \caption{Time series of  $BUL-SC39-127215$ (upper panel) and $BUL-SC39-743232$ (botton panel) stars showing the photometric contamination observed in many light curves of  the stars in the Galactic central region.}
  \label{1}
\end{figure}
Stars selected by the described process  were called {\it{Be star candidates}}. Table 1 shows also for each field of the Galactic Bulge the total number of Be star candidates. \\  
In order to obtain a subsample of the {\it{most certain Be stars}}(i.e. those stars with  similar light curves  to those of classical Be stars and  similar to those found in the Magellanic Clouds (Mennickent et al. 2002; Keller et al. 2002; Sabogal et al. 2005)), we applied a statistical filter on the stellar I magnitude distribution for each field. We first inspected visually the light curves of the stars in the five test fields mentioned before, and in field $BUL-SC5$, and selected the most certain Be stars that we used as typical Be-star like variability indicators. 
We then searched for correlations between mean magnitude ($\overline{I}$), standard deviation ($\sigma$), skewness ($S$) and  kurtosis ($K$) of the magnitude distribution for each field, and the robust estimates for these quantities $\overline{I_R}$, $\sigma_R$, $S_R$ and $K_R$, respectively,  where $S_R$ and $K_R$ are obtained by using the typical definitions of skewness ($S = (\sigma^3(N-1))^{-1} \sum_{i=1}^{N}(I_i-\overline{I})^3$) and kurtosis ($K = (\sigma^4(N-1))^{-1} \sum_{i=1}^{N}(I_i-\overline{I})^4$) but replacing $\overline{I}$ and $\sigma$ by $\overline{I_R}$ and $\sigma_R$. The aim of this step was to find a set of correlations between these parameters, whose ranges of values were the same for all the magnitude distributions in all fields, in order to select the major number of the most certain Be stars without having to inspect visually
thousands of light curves. To obtain these correlations we performed several plots of these statistical parameters and of aritmethic combinations of them (for example $K$ vs $S$, $S_R$ vs $S$, etc.) and selected only those showing very clear correlations between them.  It is important to remind that, due to the special  shapes of the light curves of the most certain Be stars sample, we expected that skewness and kurtosis of the sample would be  different  of the reminding sample of  Be star candidates, and could be usefull to find correlations that would allow us to  characterize the most certain Be stars sample in order to select these stars without  visual inspection.  Indeed we found  excellent  correlations  between  $p_{1} = \overline{I_R} / \overline{I}$, $p_{2} = \sigma_R / \sigma$, $p_{3} = S_R / S$, and $p_{4} = \vert p_{1} \times p_{2} \times p_{3}\vert$. Parameters with the best correlations and their ranges of values are shown in Table 3. By using these correlations we could recover 70 \% of the previously most certain Be stars visually selected. This means that our method will eventually allow to find a representative but not complete sample of the variable stars satisfying our color and magnitude criteria.\\
Then, we applied this filter on the total number of Be star candidates, which  selects only the stars with magnitude distributions whose  $p_{1}$, $p_{2}$, $p_{3}$ and $p_{4}$ values are in the ranges shown in Table 3. Finally we inspected visually the sample selected by this filter to obtain the subsample of most certain Be stars in the remaining fields. Table 1 shows for each field of the Galactic Bulge the total number of stars selected by this process. \\
Finally, the selection process used in this search of Be stars in  the direction of the Galactic Bulge can be summarized in the following steps: first, we searched for stars with colours and magnitudes typical for Be stars. The new sample was called Be star precandidates. From these precandidates we then selected only those variable stars with amplitudes in the typical range of variability of Be stars of the Galaxy and the Magellanic Clouds. The obtained sample was called Be star candidates. Finally we used a statistical filter followed by visual inspection to select from the Be star candidates those with  typical Be-star like variability. We called this final sample the most certain  Be star Candidates. 

\begin{table}
 \caption{Selected Fields used to define our variability-based 
 selection criteria. The epoch for the coordinates is year 2000.}
 \label{2}
 \centering
 \begin{tabular}{@{}ccccc}
  \hline
  \hline
  Field & $\alpha$(hh:mm:ss) & $\delta$(dd:mm:ss) & $l(^{\circ})$ & $b(^{\circ})$ \\
  \hline
  $BUL-SC1$ & 18:02:32.5 & 29:57:41 & 1.08 & 3.62 \\
  $BUL-SC10$ & 18:20:06.6 & 22:23:03 & 9.64 & 3.44 \\
  $BUL-SC11$ & 18:21:06.5 & 22:23:05 & 9,74 & 3.64 \\
  $BUL-SC26$ & 17:47:15.5 & 34:59:31 & 4.90 & 3.37 \\
  $BUL-SC48$ & 17:28:14.0 & 39:46:58 & 11.07 & 2.78 \\
  \hline
 \end{tabular}
\end{table}

\begin{table}
 \caption{Parameters for the best correlations.}
 \label{3}
\centering
 \begin{tabular}{@{}ccc}
  \hline
  \hline
  Name & Definition & Range of Values \\
  \hline
  $p_{1}$ & $\overline{I_R} / \overline{I}$ & $[0.9970,1.0011]$ \\
  $p_{2}$ & $\sigma_R / \sigma$ & $[0.30,1.22]$ \\
  $p_{3}$ & $S_R / S$ & $[-1.00,7.15]$ \\
  $p_{4}$ & $\vert  p_{1} \times p_{2} \times p_{3} \vert $ & $[0.29,3.82]$  \\
  \hline
 \end{tabular}
\end{table}

\subsection{Discussion of our sample of Be Star Candidates}
The final result of our search in the 48 OGLE II Galactic Bulge fields is a list of 29053 Be star candidates and 1488 most certain Be star candidates that we provide as a catalog in the digital version of this article. 
Figure 3 shows some of the light curves of the most certain Be star candidates in the direction of the Galactic bulge. HJD zero point of the data is 2450000. These light curves are typical of the selected sample, which have variable amplitudes, outbursts in some cases, small variations overlapping long-term variations, etc. 

\begin{figure}
\centering
\scalebox{.9}[.9]{\includegraphics[angle=0,width=10cm]{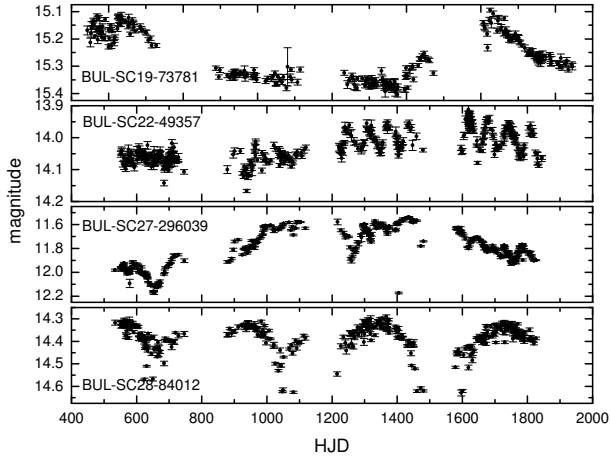}}
 \caption{Examples of light curves for the most certain Be star candidates in the direction of the Galactic bulge.}
  \label{1}
\end{figure}

Table 4 contains the catalogue of the selected Be star candidates. This table gives Field, OGLE ID of star, equatorial coordinates, $V$ magnitude, $V-I$ colour and $A_I$ amplitude (complete table is available in electronic form). The first 1488 stars in this catalogue are those called the most certain Be star candidates. In Figure 4 we show the $V$ versus $V-I$ diagram defined by the complete sample of stars. There is a clear sequence of many stars in the blue  part of the diagram (where Be stars are expected) and other sequence at the red part of the diagram that contains less stars. Because of the reddening within each field of the Galactic bulge region, it is possible that these stars are actually Be stars, although we cannot discard contamination by some red variables. Taking into account that the stellar population of the Galactic bulge is dominated by old stars and there is no evidence of massive objects, except in scattered regions and in the 1 parsec zone around the centre, where some stellar formation is detected
(Rodgers \& Harding 1989; Bertelli et al. 1995; Martins et al. 2007), only some Be star candidates could belong to these regions of the Galactic bulge.  On the other hand, since the turnoff of the Galactic bulge main sequence is around $M_V = 4$ (Reyl\'e  et al. 2004), and  based on stellar proper motions and Galaxy models, it is known that the blue sequence observed in colour-magnitude diagrams of clusters and fields in the direction of the Galactic bulge, and eventually in figure 4, is due to the young stellar population of the disk (Zoccali et al. 2001; Udalski et al. 2002; Kuijken \& Rich 2002). Hence, we conclude that many of the Be star candidates in figure 4 probably belong to the Galactic disk.\\

\begin{table*}
 \caption{Catalogue of Be Star Candidates in the Direction of the Galactic Bulge (complete table is available in electronic form). The epoch for the coordinates is year 2000.}
 \label{4}
\centering
 \begin{tabular}{@{}ccccccc}
  \hline
  \hline
  Field & ID & $\alpha$(hh:mm:ss) & $\delta$(dd:mm:ss) &  $V$ & $V-I$ & ${\bf{A_I}}$ \\
  \hline
 $BUL - SC1$ & 116088 & 18 02  04.67 & -29 48  23.7 &  15.905 &  1.211 &  0.135\\
 $BUL - SC1$ & 130042 & 18 02  13.76 & -29 45  18.8 &  15.285 &  1.05 &  0.118\\
 $BUL - SC1$ & 213107 & 18 02  16.96 & -30 16  57.0&  13.442 &  0.423 &  0.092\\
 $BUL - SC1$ & 224746 & 18 02  17.73 & -30 14  31.8 &  14.604 &  0.919 &  0.118\\
 $BUL - SC1$ & 257790 & 18 02  30.79 & -30 02  23.9 &  14.81 &  1.489 &  0.081\\
 $BUL - SC1$ & 257829 & 18 02  23.21 & -30 04  03.0 &  15.144 &  1.067 &  0.112\\
 $BUL - SC1$ & 291319 & 18 02  18.64 & -29 52  04.6 &  14.184 &  0.963 &  0.145\\
 $BUL - SC1$ & 304035 & 18 02  17.0& -29 49  04.0 &  14.275 &  0.861 &  0.104\\
 $BUL - SC1$ & 304058 & 18 02  30.7 & -29 47  35.0 &  14.316 &  1.07 &  0.181\\
 $BUL - SC1$ & 353800 & 18 02  24.25 & -29 35  50.6 &  15.093 &  0.816 &  0.129\\
 $BUL - SC1$ & 36057 & 18 02  03.41 & -30 14  18.3 &  14.634 &  1.368 &  0.129\\
 $BUL - SC1$ & 365772 & 18 02  23.09 & -29 32  19.3 &  15.502 &  0.938 &  0.130\\
 $BUL - SC1$ & 376373 & 18 02  40.19 & -30 24  12.4 &  13.999 &  0.738 &  0.316\\
 $BUL - SC1$ & 376671 & 18 02  46.84 & -30 23  42.2 &  15.813 &  1.046 &  0.176\\
 $BUL - SC1$ & 420820 & 18 02  35.11 & -30 10  05.8 &  15.328 &  0.855 &  0.103\\
 $BUL - SC1$ & 441953 & 18 02  43.56 & -30 03  28.5 &  14.995 &  0.931 &  0.115\\
 $BUL - SC1$ & 452151 & 18 02  34.63 & -29 59  51.0 &  15.606 &  0.993 &  0.118\\
 $BUL - SC1$ & 462569 & 18 02  38.07 & -29 55  10.4 &  14.002 &  0.868 &  0.097\\
 $BUL - SC1$ & 523070 & 18 02  38.2 & -29 38  58.9 &  13.454 &  1.307 &  0.406\\
  \hline
 \end{tabular}
\end{table*}

\begin{figure}
\centering
\scalebox{.9}[.9]{\includegraphics[angle=0,width=9cm]{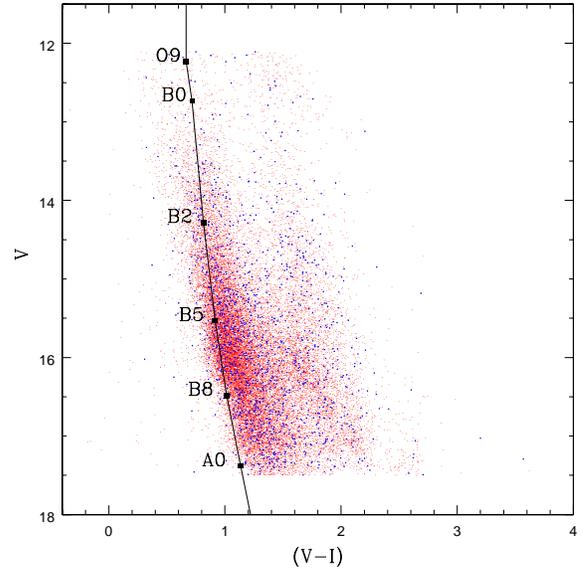}}
 \caption{$V$ vs $(V-I)$ diagram for the selected Be star candidates. The track of the main sequence (MS) (Allen 2000) is shown for reference. Apparent $V$ magnitudes for it were calculated assuming the distance modulus of the Galactic bulge (14.5 mag) and $A_V = 2.23$ (obtained by calculating the mean of $A_V$ values of  the 48 Galactic Bulge fields). Reddened colours for the MS were obtained by using $E_{(V-I)} = 1.135$, that is the mean of $E(V-I)$ values of  the 48 Galactic Bulge fields.}
  \label{2}
\end{figure}

Figure 5 shows the amplitude-magnitude  (upper graph) and the amplitude-colour (bottom graph) diagrams for the Be star candidates, where amplitude values are those  obtained  by the process described in section 3.1. It is observed from figure 5 that a large fraction of stars have amplitudes less than 0.2 magnitudes (the typical amplitude of Be stars). However, there are many stars showing amplitudes larger than 0.2.  Most of these stars have $V > 15$ mag (see amplitude-magnitude diagram) and could be Be stars with long-term variability (Hubert et al. 1997) which have amplitudes up 0.8 mag. Some of them could be also Luminous Blue Variables, which show oscillations with amplitudes of half a magnitude (Humpreys $\&$ Davidson 1994). We also observe  in the amplitude-colour diagram that stars with amplitudes greater than 0.2 magnitudes  are mostly concentrated between $(V-I)=0.8$ and $(V-I)=1.8$ mag. Based on  these aspects and on the shapes of the light curves of the selected Be star candidates, it is possible that some of these stars are variables similar to those found in the Magellanic Clouds, classified  as Type-1 to Type 3 stars (Mennickent et al. 2002). \\

\begin{figure}
\centering
\scalebox{.9}[.9]{\includegraphics[angle=0,width=9cm]{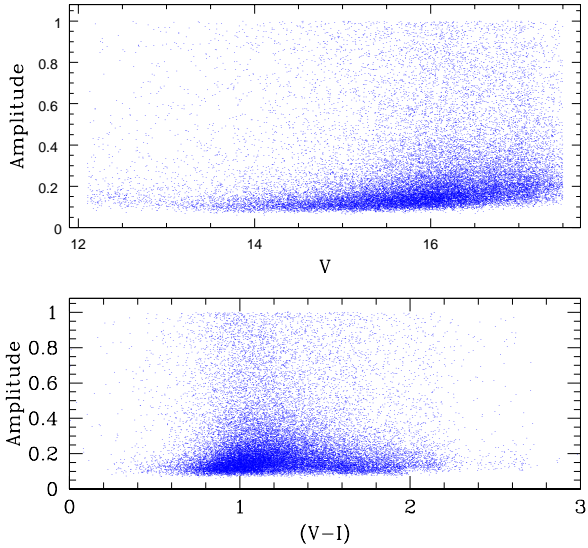}}
 \caption{I-Amplitude vs $V$ diagram (upper panel) and I-Amplitude vs $V-I$ diagram (bottom panel) for the selected Be star candidates}
  \label{3}
\end{figure}

\subsubsection{Periodic Be Star Candidates}
In order to investigate if some of the Be star candidates show short-term, mid-term or long-term variations, we searched for periodicities in their light curves.  It is important to mention that the duration and frecuency of observations of the OGLE II project were enough to detect  mid an long-term variability timescales, and also those short-term variations with periods greater than 1 day.  \\
We used for this search the analysis of variances algorithm (AOV) (Schwarzenberg-Czerny 1989). Then we used a special software provided by one of the authors  (Piertzy\'nski 2004) to graph phased light curves and confirm or reject the obtained periods. We discarded false periods and eclipsing binaries. Finally, only 198 Be star candidates (0.68 percent) turned out to be periodic variables. Table 5 presents a list of names, periods and errors of these stars (complete table is available in electronic form). These errors were calculated by measuring the halfwidth of the frequency peak in the power spectrum. Figure 6 shows phased light curves of two of the  periodic variables found. These curves and many of the phased light curves of the periodic variables found closely resemble sinusoids indicating that these stars could be binaries. It is important to mention that some of these periodic Be star candidates show several periods in the periodogram.  On the base of this fact these stars could be multiperiodic variables similar to those A and B emission-line stars studied by Mennickent et al. (2006) in the analysis of a subsample of bright type-3 stars of the SMC. We will perform a detailed study of this sample of periodic stars in a future work. For the moment we give in table 5  only the fundamental period obtained for each star.\\

\begin{figure}
\centering
 \scalebox{.9}[.9]{\includegraphics[angle=0,width=9cm]{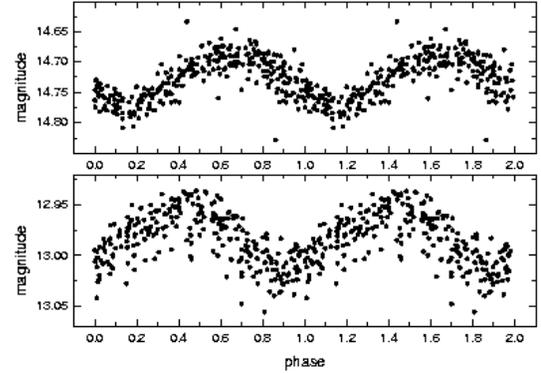}}
 \caption{{\bf{Example of light curves for two periodic variables: $BUL-SC13-204056$ with P = 7.06 days (upper panel) and $BUL-SC13-204056$, with P = 4.64 days (bottom panel).}}}
  \label{4}
\end{figure}

\begin{table}
 \caption{List of periodic Be Star Candidates in the Direction of the Galactic Bulge (complete table is available in electronic form).}
 \label{5}
 \centering
 \begin{tabular}{@{}cccc}
  \hline
  \hline
  Field & ID & Period (days) & Error (days)\\
  \hline
 $BUL -SC1$ & 365679 & 21.3 & 0.3\\
 $BUL -SC1$ & 628586 & 1.574 & 0.002\\
 $BUL - SC2$ & 178538 & 65 & 3\\
 $BUL - SC2$ & 266251 &  70 & 5\\
 $BUL - SC2$ & 517884 &  35 & 1\\
 $BUL - SC3$ & 202194 & 17.8 & 0.2\\
 $BUL - SC3$ & 328070 & 10.12 & 0.05\\
 $BUL - SC3$ & 355285 & 9.0 & 0.4\\
 $BUL - SC3$ & 355797 & 12.8 & 0.2\\
 $BUL - SC3$ & 444397 & 25.8 & 0.3\\
 $BUL - SC3$ & 455478 & 63 & 6\\
 $BUL - SC3$ & 501384 & 55 & 8\\
 $BUL - SC3$ & 525314 & 55 & 2\\
 $BUL - SC3$ & 646913 & 13.5 & 0.2\\
 $BUL - SC3$ & 657652 & 6.1 & 0.1\\
 $BUL - SC3$ & 681277 & 3.94 & 0.01\\
 $BUL - SC4$ & 120636 & 33 & 3\\
 $BUL - SC4$ & 144329 & 23.8 & 0.6\\
 $BUL - SC4$ & 155897 & 24.4 & 0.6\\
  \hline
 \end{tabular}
\end{table}

Figure 7 shows the period-amplitude diagram for the 198 periodic Be star candidates. These amplitudes were obtained by subtracting the maximum and minimum value of each light curve. These values were given by the software used to graph light curves and confirm or reject the obtained periods. \\
It is observed in figure 7 that all these stars have periods greater than 1 day. This allows us to discard the possibility that some of these stars could be $\beta$ Cepheid stars whose periods are only several hours (Stankov $\&$ Handler 2005). It is also observed in the diagram that stars have amplitudes greater than 0.06 mag, discarding the idea that they could be Slow Pulsating B stars which have amplitudes of several milimagnitudes. This fact was expected because of the selection criteria of our search, which rejected stars with amplitudes less than 0.05 mag.  The diagram in figure 7 shows the largest concentration of stars located in a range of periods between 1 and 20 days, and amplitudes between 0.06 and 0.2 mag. This is confirmed in the period and amplitude histograms shown in figure 8 (right and left panels, respectively). It shows that 30 percent of the periodic variables have periods less than 10 days and amplitudes less than 0.1 mag. This result is similar to that found for 13 Be stars in the Small Magellanic Cloud, which present short-term photometric variations with periods less than 2.5 days and amplitudes less than 0.1, superimposed to long-term variations (Martayan et al. 2007). The amplitude histogram also shows that 92 percent of the sample has amplitudes less than 0.2 mag, and only 8 percent of them have amplitudes greater than 0.2 mag. These ranges of periods and amplitudes are comparable with those of Galactic Be stars with short and mid-term photometric variability. 

\begin{figure}
\centering
 \scalebox{.9}[.9]{\includegraphics[angle=0,width=9cm]{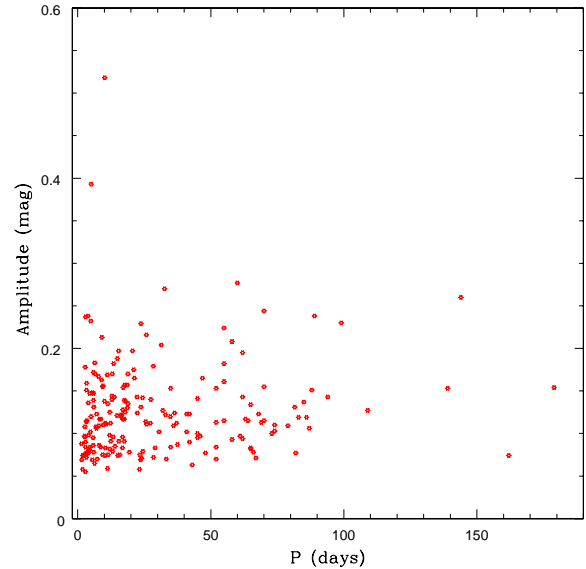}}
 \caption{Amplitude vs Period diagram for the periodic Be star candidates.}
  \label{5}
\end{figure}

\begin{figure}
\centering
\scalebox{.9}[.9]{\includegraphics[angle=0,width=9cm]{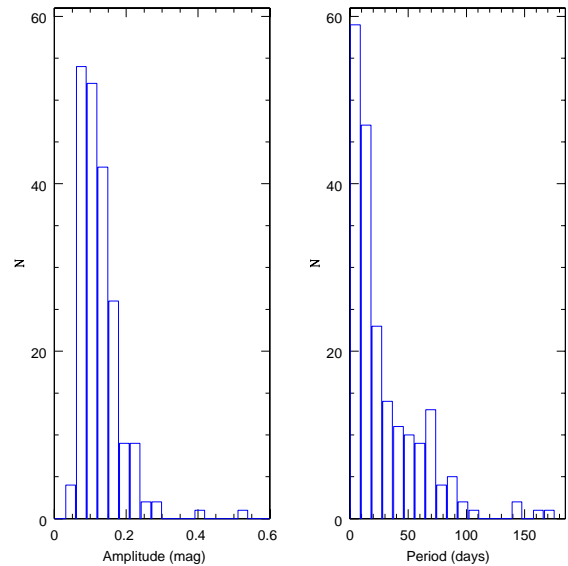}}
 \caption{Amplitude and period histogram for the periodic Be star candidates.}
  \label{6}
\end{figure}

\section{Conclusion}
In this paper we have provided a catalogue of 29053 Be star candidates, 198 of them periodic, 
in the direction of the Galactic bulge that were selected basically on photometric criteria. Most of these Be star candidates are probably members of the Galactic disk and trace the gradient of metallicity towards the Galactic centre.
They are ideal targets for future observing programs based on multiobject spectroscopy, narrow band photometry  or H$\alpha$ imaging surveys. These programs could eventually establish their true nature and broke the residual degeneracy with variable red giants in the red part of the (V-I) color distribution.

\begin{acknowledgements}
We thank to the referee, Dr Willem Jan de Witt, for his valuable suggestions and comments that improved this work. BS and AG acknowledge financial support for this work from  Programa MECESUP de Estad\'{\i}as de Estudiantes Tesistas en Centros de Excelencia en el Extranjero USA0108, and BS acknowledges support by Programa de Becas de Doctorado MECESUP UCO0209. REM acknowledges support by Fondecyt grant 1070705. WG, GP and REM acknowledge financial support for this work from the Chilean Center for Astrophysics FONDAP 15010003. \\
\end{acknowledgements}

\label{lastpage}

\end{document}